\documentclass[12pt]{article}

\usepackage{graphicx}
\usepackage{amsmath}
\usepackage{amssymb}
\usepackage{amsthm}
\numberwithin{equation}{section}

\newcommand{\Tr}{{\rm Tr}\,}
\newcommand{\grd}{{\rm grd}\,}
\newcommand{\bs}[1]{\boldsymbol{#1}}
\newcommand{\alg}[1]{\mathfrak{#1}}

\begin{document}
\baselineskip=16pt plus 0.2pt minus 0.1pt

\begin{titlepage}
\title{
\hfill{\normalsize YITP-10-39}\\
\vspace{1cm}
{\Large\bf An Algebraic Model\\
for the $\alg{su}(2|2)$ Light-Cone String Field Theory}
}
\vskip20mm
\author{
{\sc Isao Kishimoto}\thanks
{{\tt ikishimo@yukawa.kyoto-u.ac.jp}} ${}^1$\;
\quad and \quad
{\sc Sanefumi Moriyama}\thanks
{{\tt moriyama@math.nagoya-u.ac.jp}} ${}^{2,1}$\\[15pt]
${}^1${\it Yukawa Institute for Theoretical Physics,}\\
{\it Kyoto University, Kyoto 606-8502, Japan}
\medskip\\
${}^2${\it Kobayashi-Maskawa Institute 
\& Graduate School of Mathematics,}\\
{\it Nagoya University, Nagoya 464-8602, Japan}
}
\date{\normalsize May, 2010}
\maketitle
\thispagestyle{empty}
\begin{abstract}
\normalsize
We investigate algebraic structure of light-cone string field theory
which respects the spacetime supersymmetry $\alg{su}(2|2)$.
Extracting building blocks from the explicit oscillator expressions of
light-cone superstring field theory on the pp-wave and the flat
background, we propose a model for more generic backgrounds by giving
an algebraic procedure to construct supercharges and Hamiltonian
without referring to explicit oscillator expressions.
Based on this strategy, we systematically construct interaction terms
of light-cone string field theory on the pp-wave background and some
examples of its generalization.
We also study the supergravity limit of the interaction terms on the
pp-wave background.
\end{abstract}
\end{titlepage}

\section{Introduction}

There is no overemphasizing the importance of the superalgebra
$\alg{su}(2|2)=\alg{psu}(2|2)\ltimes{\mathbb R}$ in the recent
progress of AdS/CFT correspondence.
The first work which was aware of its importance is probably
\cite{Bsu23}, which shows that the expressions of various
superconformal charges of super Yang-Mills theory are strongly
constrained in the $\alg{su}(2|3)$ subsector.
The appearance of the superalgebra $\alg{su}(2|3)$ was further
clarified by the reason that it produces a similar stabilizer
subalgebra $[\alg{psu}(2|2)]\ltimes{\mathbb R}$ as the whole symmetry
superalgebra $\alg{psu}(2,2|4)$ produces the stabilizer subalgebra
$[\alg{psu}(2|2)]^2\ltimes{\mathbb R}$ after fixing the vacuum
\cite{Bthesis}.
Also excitingly, it was discovered that this powerful superalgebra
successfully fixes the scattering matrix up to an overall factor
\cite{BSmat}, which satisfies the integrable Yang-Baxter equation and
the Yangian symmetry.
Hence, it is tantalizing to generalize the analysis for single objects
(trace operators or strings) to their interactions.
As on the pp-wave background, the first step to study the string
interactions is to construct a light-cone string field
theory\footnote{See \cite{CJJK} for an interesting attempt using the
coherent states.}.

Another important related research is to classify the geometrical
background preserving half of the supersymmetries \cite{LLM}.
Although the superalgebra was not explicitly mentioned in the original
work, it is clear from the recent development that this work amounts
to deforming the $AdS_5\times S^5$ background under the requirement
that the background preserves the subisometry
$[\alg{psu}(2|2)]^2\ltimes{\mathbb R}$.
The deformed background was named bubbling geometries after their
construction.
Our above direction can be reexpressed as construction of light-cone
string field theory on the bubbling geometries.

Compared with AdS/CFT correspondence, construction of light-cone
string field theory has a much longer history.
The three-string interaction vertex of light-cone superstring field
theory on the flat spacetime was constructed in eighties \cite{GSB},
by generalizing the bosonic light-cone string field theory \cite{KK}
and the extended supergravity \cite{GS}.
The basic building blocks are {\it the kinematical overlap}
$|V\rangle$ constructed from the local fermionic (and bosonic)
momentum conservation on the worldsheet
\begin{align}
\bigl(\lambda_{(1)}(\sigma)
+\lambda_{(2)}(\sigma)
+\lambda_{(3)}(\sigma)\bigr)
|V\rangle=0,\quad
\bigl(p_{(1)}(\sigma)
+p_{(2)}(\sigma)
+p_{(3)}(\sigma)\bigr)
|V\rangle=0,
\label{overlap}
\end{align}
with $\lambda_{(i)}(\sigma)$ (and $p_{(i)}(\sigma)$) denoting the fermionic
(and bosonic) momentum of the $i$-th string, and {\it a real fermionic
operator} $Y$ which commutes with the local momentum conservation and
is interpreted as the renormalized fermionic momentum at the
interaction point $\sigma_{\rm I}$ of the light-cone three-string
kinematical overlap,
\begin{align}
\lambda_{(i)}(\sigma)|V\rangle
&\sim\frac{1}{\sqrt{\sigma-\sigma_{\rm I}}}Y|V\rangle,
\label{renorm}
\end{align}
with {\it its supersymmetric partners} $X,\widetilde X$ defined as
\begin{align}
\{q,Y\}=X,\quad\{\widetilde q,Y\}=\widetilde X,
\end{align}
which can also be interpreted as the holomorphic/antiholomorphic part
of the bosonic momentum at the interaction point $\sigma_{\rm I}$.
The expressions of various charges at the interacting level are then
determined from the supersymmetry algebra.

More recently, the light-cone superstring field theory on the pp-wave
background was constructed \cite{SVPS} in the development of AdS/CFT
correspondence by simply generalizing that on the flat spacetime,
which enables the match between Hamiltonian of string theory and the
dilatation of super Yang-Mills theory at the interacting level
\cite{GMP}.
Subsequently, it was pointed out that another linear combination
proportional to the action of the free charges on the kinematical
overlap $|V\rangle$ can also solve the algebra as well \cite{dVPPRT}.
In the context of AdS/CFT correspondence it was then proposed \cite{DY}
that we have to take the original part coming from the flat spacetime
and the part proportional to the free action with equal weight in
order to match the correlation functions\footnote{For an interesting
computation on the match between Hamiltonian and the dilatation using
\cite{DY}, see \cite{GORSY}.}.
However, so far there is no concrete reasoning a priori (without
referring to AdS/CFT correspondence) why we have to take the two parts
with equal rate\footnote{See \cite{LR} for various stimulating
arguments in this direction.}.

In this paper we would like to revisit the construction of light-cone
string field theory on the pp-wave background and take a step towards
constructing light-cone string field theory on a background with the
isometry $\alg{su}(2|2)$.
In revisiting the construction on the pp-wave background, we have made
several clarifications.
Especially, we present a systematic analysis for terms consistent with
the superalgebra $\alg{su}(2|2)$ and determine some of the
coefficients of these terms from the supergravity limit.
After these clarifications, we propose a strategy for a general
bubbling geometry and construct some explicit examples of
generalization.
We hope that our attempts in construction will finally result in the
success to match between Hamiltonian of string theory on the bubbling
geometries with the isometry $\alg{su}(2|2)$ and anomalous dimension
of multi-trace operators as in the $\alg{su}(2|2)$ spin chain.

The construction we adopt in this paper is possible due to several
observations.
The first observation is that the isometry algebra of the pp-wave
background contains $[\alg{psu}(2|2)]^2\ltimes{\mathbb R}$ as its
subalgebra and the subalgebra $[\alg{psu}(2|2)]^2\ltimes{\mathbb R}$
acts on the other generators of the pp-wave isometry as an outer
automorphism.
This fact implies the central role of the subalgebra
$[\alg{psu}(2|2)]^2\ltimes{\mathbb R}$.
Namely, the stabilizer subalgebra
$[\alg{psu}(2|2)]^2\ltimes{\mathbb R}$ is crucial and ubiquitous while
the extra enhanced symmetry only appears accidentally in the pp-wave
limit.
Then, in retrospect to the construction of light-cone superstring
field theory on the pp-wave background in the
$\alg{so}(4)\times\alg{so}(4)$ formalism \cite{P}, we shall notice in
the following section that only the subalgebra
$[\alg{psu}(2|2)]^2\ltimes{\mathbb R}$ plays an essential role in the
construction.

The second observation is the similarity in the situation between the
construction of light-cone superstring field theory on the flat
spacetime using the super Poincare algebra and that of the spin chain
in modeling the computation of the anomalous dimension in super
Yang-Mills theory using the superalgebra
$\alg{psu}(2|2)\ltimes{\mathbb R}^3$.
In the case of the light-cone superstring field theory, the
conventionally written super Poincare algebra on the flat spacetime is
\begin{align}
\{Q^A,Q^B\}=\{{\widetilde Q}^A,{\widetilde Q}^B\}=2\delta^{AB}H,\quad
\{Q^A,{\widetilde Q}^B\}=0.
\label{flat}
\end{align}
However, the charges never satisfy these anticommutation relations
literally.
In general, at the first sight, the Hamiltonian computed from
left-moving supercharges $\{Q^A,Q^B\}$ and that computed from
right-moving ones $\{{\widetilde Q}^A,{\widetilde Q}^B\}$ do not
coincide and the anticommutator $\{Q^A,{\widetilde Q}^B\}$ does not
vanish.
The algebra \eqref{flat} holds only after we impose the level-matching
condition.
The situation was more clearly formulated in the case of the AdS/CFT
spin chain model \cite{BSmat}.
Although the on-shell symmetry is $\alg{psu}(2|2)\ltimes{\mathbb R}$,
to take care of the off-shell states we have to consider the centrally
extended superalgebra $\alg{psu}(2|2)\ltimes{\mathbb R}^3$ instead in
the spin chain model.
We shall exploit this centrally extended off-shell formalism also for the
light-cone superstring field theory.
Namely, instead of the original algebra
$\alg{psu}(2|2)\ltimes{\mathbb R}$, we shall choose the centrally
extended one $\alg{psu}(2|2)\ltimes{\mathbb R}^3$ as the off-shell
symmetry of the light-cone superstring field theory.

A natural question which may arise here is how it can be possible to
construct the light-cone string field theory if we have not solved the
worldsheet theory.
In the typical string field theory, the string worldsheet theory is
solved explicitly by string oscillators and all of the building blocks
are expressed explicitly in terms of these oscillators.
Here, without the explicit solutions to the worldsheet
theory\footnote{The spirit is somehow similar to \cite{KS}.}, we
simply assume the building blocks to be some abstract quantities
subject to suitable algebraic relations and do not refer to its
explicit form.
We hope that explicit expressions for the building blocks will be
found in the near future.

In the following section we shall revisit light-cone superstring field
theory on the pp-wave background.
After revisiting the construction on the pp-wave background, in the
subsequent section we shall proceed to propose a strategy for a more
general background with the isometry $\alg{su}(2|2)$.

\section{PP-wave light-cone string field theory revisited}
In this section, we shall revisit the three-string interaction vertex
of light-cone string field theory on the pp-wave background.
Most of the computations here essentially appeared in the previous
works \cite{P,LR}.
However, we would like to emphasize the following improvements.
\begin{itemize}
\item
After hiding the explicit oscillator expressions of various building
blocks, the algebraic aspect of the construction is much clearer.
In particular, from our following analysis, it is clear that most
expressions of the three-string interaction vertex are determined
purely from the algebraic consistency.
The generalization in the subsequent section is based heavily on the
analysis here.
\item
We do not need separate analysis for quantities with various powers,
as in \cite{P}.
Stimulated by the hyperbolic expression in \cite{KM}, we shall present
a systematic manipulation for various quantities.
The formulas in the appendix lead us directly to the final expression.
\item
The dependence on the worldsheet coordinates had to be introduced in
\cite{GSB,P} to show the uncorrelation between the left-moving and
right-moving supercharges $\{Q,\widetilde Q\}=0$.
Here in subsection 2.2, we shall prove the same equation algebraically
without referring to the dependence on the worldsheet coordinates.
\item
Although it was noticed \cite{dVPPRT,LR} that action of free charges
on the (dressed) kinematical overlaps $|V\rangle$ satisfies the
algebra as well, so far there is no explicit analysis that this is the
only alternative we have to take into account.
Our systematic analysis in subsection 2.3 clarifies all the
alternatives.
\item
The requirement of the abelian duality symmetry in the supergravity
limit determines some of the coefficients which is consistent with the
previous work \cite{DY}.
(See also \cite{LR}.)
\end{itemize}

\subsection{The building blocks}
Let us start with the superalgebra 
$\alg{psu}(2|2)\ltimes{\mathbb R}^3$ \cite{BSmat},
\begin{align}
\{{\mathfrak Q}^\alpha{}_a,{\mathfrak Q}^\beta{}_b\}
&=\epsilon^{\alpha\beta}\epsilon_{ab}{\mathfrak P},\nonumber\\
\{{\mathfrak S}^a{}_\alpha,{\mathfrak S}^b{}_\beta\}
&=\epsilon^{ab}\epsilon_{\alpha\beta}{\mathfrak K},\nonumber\\
\{{\mathfrak Q}^\alpha{}_a,{\mathfrak S}^b{}_\beta\}
&=\delta^b_a{\mathfrak L}^\alpha{}_\beta
+\delta^\alpha_\beta{\mathfrak R}^b{}_a
+\delta^b_a\delta^\alpha_\beta{\mathfrak C}.
\end{align}
If we redefine various generators by ($\eta=e^{\pi i/4}$)
\begin{align}
&\mathcal R{}^a{}_b=-i{\mathfrak R}^a{}_b,
\quad
\mathcal L{}^\alpha{}_\beta=i{\mathfrak L}^\alpha{}_\beta,
\quad
\mathcal H
={\mathfrak C}+i\frac{{\mathfrak P}-{\mathfrak K}}{2},
\quad
\mathcal{\widetilde H}
={\mathfrak C}-i\frac{{\mathfrak P}-{\mathfrak K}}{2},
\quad
\mathcal N=\frac{{\mathfrak P}+{\mathfrak K}}{2},
\nonumber\\
&\mathcal Q{}^\alpha{}_a
=\frac{1}{\sqrt{2}}(\eta{\mathfrak Q}^\alpha{}_a
+\eta^*\epsilon^{\alpha\beta}\epsilon_{ab}
{\mathfrak S}^b{}_\beta),
\quad
\mathcal{\widetilde Q}{}^\alpha{}_a
=\frac{1}{\sqrt{2}}(\eta^*{\mathfrak Q}^\alpha{}_a
+\eta\epsilon^{\alpha\beta}\epsilon_{ab}
{\mathfrak S}^b{}_\beta),
\end{align}
we can rewrite the algebra into the following expression,
\begin{align}
\{\mathcal Q{}^\alpha{}_a,\mathcal Q{}^\beta{}_b\}
&=\epsilon^{\alpha\beta}\epsilon_{ab}\mathcal H,\nonumber\\
\{\mathcal{\widetilde Q}{}^\alpha{}_a,\mathcal{\widetilde Q}{}^\beta{}_b\}
&=\epsilon^{\alpha\beta}\epsilon_{ab}\mathcal{\widetilde H},\nonumber\\
\{\mathcal Q{}^\alpha{}_a,\mathcal{\widetilde Q}{}^\beta{}_b\}
&=\epsilon^{\alpha\beta}\epsilon_{ac}\mathcal R{}^c{}_b
+\epsilon_{ab}\mathcal L{}^\alpha{}_\gamma\epsilon^{\gamma\beta}
+\epsilon^{\alpha\beta}\epsilon_{ab}\mathcal N.
\label{alg}
\end{align}
Let us expand each generator $\mathcal J$ with respect to the string
coupling constant $g_s$, $\mathcal J=j+g_sJ+\cdots$, where the first
term is the free generator $j$ acting on each of three strings and the
second one is their interaction $J$, which is conventionally expressed
in the ket form $|J\rangle$.
In terms of these generators, the superalgebra becomes
\begin{align}
q^\alpha{}_a|Q^\beta{}_b\rangle
+q^\beta{}_b|Q^\alpha{}_a\rangle
&=\epsilon^{\alpha\beta}\epsilon_{ab}
|H\rangle,\nonumber\\
{\widetilde q}^\alpha{}_a|{\widetilde Q}^\beta{}_b\rangle
+{\widetilde q}^\beta{}_b|{\widetilde Q}^\alpha{}_a\rangle
&=\epsilon^{\alpha\beta}\epsilon_{ab}
|{\widetilde H}\rangle,\nonumber\\
q^\alpha{}_a|{\widetilde Q}^\beta{}_b\rangle
+{\widetilde q}^\beta{}_b|Q^\alpha{}_a\rangle
&=\epsilon^{\alpha\beta}\epsilon_{ab}|N\rangle,
\label{superalg}
\end{align}
where we have assumed the bosonic generators
${\mathcal L}^\alpha{}_\beta$ and ${\mathcal R}^a{}_b$ to be
kinematical ones without receiving any interacting corrections just as
the canonical role these generators played in the $\alg{su}(2|2)$ spin
chain model \cite{Bsu23,BSmat}.

Now let us turn to the construction of the light-cone string field
theory on the pp-wave background \cite{P}.
In the following of this subsection, we shall encounter several
assumptions which cannot be explained only from the symmetry algebra
$\alg{psu}(2|2)\ltimes{\mathbb R}^3$.
The justification for these assumptions is, of course, due to the
explicit computations on the pp-wave background, where the worldsheet
theory is exactly solvable.
After the discussions in this section, it will be clear which part has
to be deformed when we extend the background from the pp-wave one to a
more general one with the isometry
$\alg{psu}(2|2)\ltimes{\mathbb R}^3$.

In describing the three-string interaction of light-cone string field
theory, the main building blocks are the kinematical overlap among
three strings $|V\rangle$ and the renormalized fermionic momentum $Y$
as defined in \eqref{overlap} and \eqref{renorm}.
The definition of $|V\rangle$ \eqref{overlap} refers only to the
conservation law of the worldsheet momentum, which should remain valid
in a general background.
The singular behavior of \eqref{renorm} also originates from the
worldsheet instead of the spacetime and is interpreted as a
two-dimensional operator product expansion \cite{KMT,KM}.
Hence, it is natural to expect that the definition of $Y$ also
remains valid in a general background.
The indices of $Y$ are determined naturally as being the (fermionic)
spacetime momentum.

A strong assumption is that the renormalized momenta transform under
the supersymmetry linearly as\footnote{The extra minus sign in the
first two lines can be regarded as the minus sign appearing in
$\epsilon^{\alpha\beta}\epsilon_{\beta\gamma}=-\delta^\alpha_\gamma$,
$\epsilon^{ab}\epsilon_{bc}=-\delta^a_c$, while the meaning of the
overall coefficients in the last line will be clear later in this
subsection.}
\begin{align}
\{q^\alpha{}_a,Y^\beta{}_{\dot b}\}&
=-x\epsilon^{\beta\alpha}(\epsilon X)_{a\dot b},&
\{q^\alpha{}_a,Y'^b{}_{\dot\beta}\}&
=x'\delta^b_aX'^\alpha{}_{\dot\beta},\nonumber\\
[q^\alpha{}_a,X{}^b{}_{\dot b}]&
=w\delta^b_aW^\alpha{}_{\dot b},&
[q^\alpha{}_a,X'^\beta{}_{\dot\beta}]&
=-w'\epsilon^{\beta\alpha}(\epsilon W')_{a\dot\beta},\nonumber\\
[q^\alpha{}_a,{\widetilde X}{}^b{}_{\dot b}]&
=\frac{i}{2}y\delta^b_aY^\alpha{}_{\dot b},&
[q^\alpha{}_a,{\widetilde X}{}'^\beta{}_{\dot\beta}]&
=\frac{i}{2}y'\epsilon^{\beta\alpha}(\epsilon Y')_{a\dot\beta},
\label{CR}
\end{align}
with $(\epsilon X)_{a\dot b}=\epsilon_{ab}X^b{}_{\dot b}$,
$(\epsilon W')_{a\dot\beta}=\epsilon_{ab}W'^b{}_{\dot\beta}$,
$(\epsilon Y')_{a\dot\beta}=\epsilon_{ab}Y'^b{}_{\dot\beta}$.
In fact, we can regard the first two equations as the definition of
the quantities $X$, $X'$, $W$, $W'$ on the right-hand side.
Therefore, the only assumption we have actually made is the last one.
If we switch the indices by contracting with $\epsilon$ such as
$Y^{\dot b}{}_\beta
=\epsilon_{\beta\alpha}Y^\alpha{}_{\dot a}\epsilon^{\dot a\dot b}$,
$Y'{}^{\dot\beta}{}_b
=\epsilon_{ba}Y'{}^a{}_{\dot\alpha}\epsilon^{\dot\alpha\dot\beta}$,
the transformation rule is rewritten as
\begin{align}
\{q^\alpha{}_a,Y^{\dot b}{}_\beta\}&
=xX^{\dot b}{}_a\delta^\alpha_\beta,&
\{q^\alpha{}_a,Y'^{\dot\beta}{}_b\}&
=-x'(X'\epsilon)^{{\dot\beta}\alpha}\epsilon_{ab},
\nonumber\\
[q^\alpha{}_a,X^{\dot b}{}_b]&
=-w(W\epsilon)^{{\dot b}\alpha}\epsilon_{ab},&
[q^\alpha{}_a,X'^{\dot\beta}{}_\beta]&
=w'W'{}^{\dot\beta}{}_a\delta^\alpha_\beta,
\nonumber\\
[q^\alpha{}_a,{\widetilde X}{}^{\dot b}{}_b]&
=-\frac{i}{2}y(Y\epsilon)^{{\dot b}\alpha}\epsilon_{ab},&
[q^\alpha{}_a,{\widetilde X}{}'^{\dot\beta}{}_\beta]&
=-\frac{i}{2}y'Y'{}^{\dot\beta}{}_a\delta^\alpha_\beta.
\end{align}
Besides, we shall require another strong assumption. 
We assume that the action of the supersymmetry charge on the
kinematical overlap $|V\rangle$ is given as
\begin{align}
q^\alpha{}_a|V\rangle
&=\frac{i}{2}[v(YX)^\alpha{}_a+v'(X'Y')^\alpha{}_a]|V\rangle.
\label{qV}
\end{align}
We also assume the supersymmetry generator $\widetilde q^\alpha{}_a$
acts similarly as $q^\alpha{}_a$, with all of the quantities replaced
by those with tildes (except $Y$ and $Y'$) and complex conjugation
taken for complex numbers.

{}From the consistency with the first line of \eqref{alg} with the
full generators $\mathcal J$ replaced by the free one $j$, we obtain
the following relations
\begin{align}
[h,Y^\alpha{}_{\dot a}]&=xwW^\alpha{}_{\dot a},&
[h,Y'^a{}_{\dot\alpha}]&=x'w'W'^a{}_{\dot\alpha},
\label{eq:hY=W}
\end{align}
as well as
\begin{align}
h|V\rangle
=\frac{i}{2}(vx\Tr X^2+v'x'\Tr X'^2-vw\Tr YW+v'w'\Tr W'Y')|V\rangle,
\end{align}
while the consistency with the third line of \eqref{alg} requires
\begin{align}
\frac{\widetilde x}{x}
=\frac{\widetilde y}{y}
=\frac{\widetilde v}{v},\quad
\frac{{\widetilde x}'}{x'}
=\frac{{\widetilde y}'}{y'}
=\frac{{\widetilde v}'}{v'},\quad
x{\widetilde y}+{\widetilde x}y=2,\quad
x'{\widetilde y}'+{\widetilde x}'y'=2.
\label{eq:quv_consistency}
\end{align}

Looking back to the definition of \eqref{CR}, it is easily seen that
$x$ (and $w$) fix the relative ratio between $Y$ and $X$ (and $X$ and
$W$, respectively).
Hence, we can set them to $1$ by the redefinition of $X$ and $W$.
Similarly, we can also set $\widetilde x$ and $\widetilde w$ to 1 by
the redefinition of $\widetilde X$ and $\widetilde W$.
In addition, as seen from \eqref{qV}, $v$ fixes the overall factor of
$Y$ and we also set it to $1$.
Subsequently, from \eqref{eq:quv_consistency} we find $\widetilde v$,
$y$ and $\widetilde y$ are determined to be $1$.
All these arguments hold also for the variables with primes.
Therefore, hereafter without loss of generality, we shall take
\begin{align}
x=x'=w=w'=y=y'=v=v'
=\widetilde x=\widetilde x'=\widetilde w=\widetilde w'
=\widetilde y=\widetilde y'=\widetilde v=\widetilde v'=1.
\end{align}

\subsection{Anti-chiral terms}
Now let us take the interaction terms of the supersymmetry charges to
be the most general ones with one bosonic momentum inserted (where the
opposite chirality terms with $\widetilde X$ and $\widetilde X{}'$
replaced by $X$ and $X'$ are postponed to the next subsection),
\begin{align}
|Q^\alpha{}_a\rangle&
=\Bigl\{\sum_{n,m}q_{nm}(Y^n\widetilde XY'{}^m)^\alpha{}_a
+\sum_{n,m}q'_{mn}(Y^m\widetilde X{}'Y'{}^n)^\alpha{}_a\Bigr\}
|V\rangle,
\label{eq:onshell}
\end{align}
with $n$ denoting odd numbers $1,3$ and $m$ denoting even numbers
$0,2,4$.
Then, as we study in the following, we can impose a strong constraint
on the coefficients $q_{nm}$ and $q'_{mn}$ from the supersymmetry
algebra.
Although most of the computations essentially appeared in \cite{P},
let us repeat the computations here in a much more systematic way.

Before starting the analysis, let us note several properties which
will be necessary hereafter.
The first important property is the symmetry of the products
\begin{align}
&(Y^4)^\alpha{}_\beta=\delta^\alpha_\beta\frac{1}{2}\Tr Y^4,\quad
(Y^4)^{\dot a}{}_{\dot b}=-\delta^{\dot a}_{\dot b}\frac{1}{2}\Tr Y^4,\quad
(X^2)^b{}_a=\delta^b_a\frac{1}{2}\Tr X^2,\nonumber\\
&\Tr Y^2=0,\quad
(Y^2)^\alpha{}_\beta(Y^2)^{\dot a}{}_{\dot b}=0,
\end{align}
where we choose the convention that ${\rm Tr}$ is taken over the
undotted indices: $\Tr Y^4=(Y^4)^\alpha{}_\alpha$, $\Tr X^2=(X^2)^a{}_a$.
Secondly, the following formulas also play important roles in the
calculation.
\begin{align}
&(V\epsilon)^{\alpha\beta}-(V\epsilon)^{\beta\alpha}
=\epsilon^{\alpha\beta}\Tr V,\qquad
(\epsilon W)_{ab}-(\epsilon W)_{ba}
=\epsilon_{ab}\Tr W,\nonumber\\
&V^\alpha{}_aW^\beta{}_b-V^\beta{}_aW^\alpha{}_b
=-\epsilon^{\alpha\beta}(\epsilon \bar VW)_{ab},\quad
V^\alpha{}_aW^\beta{}_b-V^\alpha{}_bW^\beta{}_a
=-\epsilon_{ab}(V\bar W\epsilon)^{\alpha\beta},\nonumber\\
&V^\alpha{}_aW^\beta{}_b+V^\beta{}_bW^\alpha{}_a
-V^\beta{}_aW^\alpha{}_b-V^\alpha{}_bW^\beta{}_a
=-\epsilon^{\alpha\beta}\epsilon_{ab}\Tr V\bar W.
\end{align}
Note that $V$, $W$ are some general products, with $\bar V$, $\bar W$
denoting $\bar V=\epsilon V^{\rm T}\epsilon$,
$\bar W=\epsilon W^{\rm T}\epsilon$,
where in the transposes $V^{\rm T}$, $W^{\rm T}$ the signs
coming from exchanges among fermions have to be taken into account.

Now we shall turn to the main formulas.
The following simple formulas hold for the action of the free
supercharges $q^\alpha{}_a$ on the prefactors 
$(Y^n\widetilde XY'^m)^\beta{}_b$,
$(Y^m\widetilde X'Y'^n)^\beta{}_b$ ($n=1,3$, $m=0,2,4$) and the
kinematical overlap $|V\rangle$.
(For the derivation of these formulas, see appendix A.)
\begin{align}
\bigl\{q^\alpha{}_a,(Y^n\widetilde XY'^m)^\beta{}_b\bigr\}
&=-\frac{n}{2}
\bigl[\epsilon^{\beta\alpha}
(\epsilon XY^{n-1}\widetilde XY'^m)_{ab}
+(Y^{n-1}\epsilon)^{\beta\alpha}
(\epsilon X\widetilde XY'^m)_{ab}\bigr]\nonumber\\
&\hspace{-32mm}
+\frac{i}{2}(Y^{n+1}\epsilon)^{\beta\alpha}(\epsilon Y'^m)_{ab}
-\frac{m}{2}
\bigl[(Y^n\widetilde X)^\beta{}_a
(X'Y'^{m-1})^\alpha{}_b
+(Y^n\widetilde XY'^{m-1}X'\epsilon)^{\beta\alpha}
\epsilon_{ab}\bigr],\label{qnm}\\
(YX)^\alpha{}_a(Y^n\widetilde XY'^m)^\beta{}_b
&=-\frac{1}{n+1}
\bigl[\epsilon^{\beta\alpha}
(\epsilon XY^{n+1}\widetilde XY'^m)_{ab}
-(Y^{n+1}\epsilon)^{\beta\alpha}
(\epsilon X\widetilde XY'^m)_{ab}\bigr],\label{YXnm}\\
(X'Y')^\alpha{}_a(Y^n\widetilde XY'^m)^\beta{}_b
&=-\frac{1}{m+1}
\bigl[(Y^n\widetilde X)^\beta{}_a
(X'Y'^{m+1})^\alpha{}_b
-(Y^n\widetilde XY'^{m+1}X'\epsilon)^{\beta\alpha}
\epsilon_{ab}\bigr],\label{XYnm}\\
\bigl\{q^\alpha{}_a,(Y^m\widetilde X'Y'^n)^\beta{}_b\bigr\}
&=-\frac{m}{2}
\bigl[\epsilon^{\beta\alpha}
(\epsilon XY^{m-1}\widetilde X'Y'^n)_{ab}
+(Y^{m-1}X)^\beta{}_a
(\widetilde X'Y'^n)^\alpha{}_b\bigr]\nonumber\\
&\hspace{-32mm}
+\frac{i}{2}(Y^m\epsilon)^{\beta\alpha}(\epsilon Y'^{n+1})_{ab}
-\frac{n}{2}
\bigl[(Y^m\widetilde X'X'\epsilon)^{\beta\alpha}
(\epsilon Y'^{n-1})_{ab}
+(Y^m\widetilde X'Y'^{n-1}X'\epsilon)^{\beta\alpha}
\epsilon_{ab}\bigr],\label{qmn}\\
(YX)^\alpha{}_a(Y^m\widetilde X'Y'^n)^\beta{}_b
&=-\frac{1}{m+1}
\bigl[\epsilon^{\beta\alpha}
(\epsilon XY^{m+1}\widetilde X'Y'^n)_{ab}
-(Y^{m+1}X)^\beta{}_a
(\widetilde X'Y'^n)^\alpha{}_b\bigr],\label{YXmn}\\
(X'Y')^\alpha{}_a(Y^m\widetilde X'Y'^n)^\beta{}_b
&=-\frac{1}{n+1}
\bigl[(Y^m\widetilde X'X'\epsilon)^{\beta\alpha}
(\epsilon Y'^{n+1})_{ab}
-(Y^m\widetilde X'Y'^{n+1}X'\epsilon)^{\beta\alpha}
\epsilon_{ab}\bigr].\label{XYmn}
\end{align}
Note that although the formulas are originally defined for $n=1,3$,
\eqref{qnm} and \eqref{qmn} also holds for $n=5$.
Consistency with the first equation of \eqref{superalg} requires the
coefficients in the supersymmetry charge to satisfy
\begin{align}
-3q_{3m}+\frac{i}{2}q_{1m}&=0,&
-4q_{n4}-\frac{i}{3}q_{n2}&=0,&
-2q_{n2}-\frac{i}{1}q_{n0}&=0,\nonumber\\
-3q'_{m3}-\frac{i}{2}q'_{m1}&=0,&
-4q'_{4n}+\frac{i}{3}q'_{2n}&=0,&
-2q'_{2n}+\frac{i}{1}q'_{0n}&=0,
\label{coeff}
\end{align}
which can be solved by
\begin{align}
q_{nm}=q\frac{\eta^n\eta^{*m}}{n!m!},\quad
q'_{mn}=q'\frac{\eta^m\eta^{*n}}{m!n!}.
\end{align}
Consequently, the supersymmetry charge and the Hamiltonian are given
by
\begin{align}
|Q^\alpha{}_a\rangle&=
\Bigl\{q\bigl[(\sinh{\overline Y}){\widetilde X}
(\cosh{\overline Y}{}')\bigr]^\alpha{}_a
+q'\bigl[(\cosh{\overline Y}){\widetilde X'}
(\sinh{\overline Y}{}')\bigr]^\alpha{}_a\Bigr\}|V\rangle,
\nonumber\\
|H\rangle&=\Bigl\{
b\bigl[\Tr Y{}^4-\Tr Y{}'^4\bigr]
\nonumber\\
&\quad+q
\bigl[\eta\Tr X\cosh{\overline Y}
{\widetilde X}\cosh{\overline Y}{}'
+\eta^*\Tr\sinh{\overline Y}{\widetilde X}
\sinh{\overline Y}{}'X{}'\bigr]
\nonumber\\
&\quad+q'
\bigr[\eta^*\Tr\cosh{\overline Y}{\widetilde X}'
\cosh{\overline Y}{}'X{}'
+\eta\Tr X\sinh{\overline Y}
{\widetilde X}{}'\sinh{\overline Y}{}'\bigr]
\Bigr\}
|V\rangle,
\label{exp}
\end{align}
with 
${\overline Y}=Y\eta$, ${\overline Y}{}'=Y'\eta^*$ and $b$ defined as
\begin{align}
b=\frac{q\eta}{12}=\frac{q'\eta^*}{12}.
\end{align}
Similarly, the second equation of \eqref{superalg} determines
$|{\widetilde Q}^\alpha{}_a\rangle$ and $|{\widetilde H}\rangle$.

The expression of supersymmetry charge and Hamiltonian in \eqref{exp}
is one of the most famous results from the pp-wave light-cone string
field theory \cite{P}, though it has never been written down in this
succinct form using the hyperbolic function.
The hyperbolic expression of the light-cone string field theory
basically originates from our experience on the flat spacetime with
$\alg{so}(8)$ symmetry \cite{KM} (with slightly change of notation):
\begin{align}
|Q^{\dot a}\rangle&=\sqrt{-\alpha}
[\sinh Y\hspace{-3mm}\hbox{/}\hspace{1mm}]^{\dot ai}
\widetilde X^i|V\rangle,\nonumber\\
|\widetilde Q^{\dot a}\rangle&=i\sqrt{-\alpha}X^i
[\sinh Y\hspace{-3mm}\hbox{/}\hspace{1mm}]^{i\dot a}
|V\rangle,\nonumber\\
|H\rangle&=X^i
[\cosh Y\hspace{-3mm}\hbox{/}\hspace{1mm}]^{ij}
\widetilde X^j|V\rangle.
\end{align}
Decomposition from the $\alg{so}(8)$-invariant expression \cite{KM}
into the current $\alg{so}(4)\times\alg{so}(4)$-invariant expression
is reminiscent of the summation formulas of the hyperbolic functions:
$\cosh(\chi_1+\chi_2)=\cosh\chi_1\cosh\chi_2+\sinh\chi_1\sinh\chi_2$, 
$\sinh(\chi_1+\chi_2)=\sinh\chi_1\cosh\chi_2+\cosh\chi_1\sinh\chi_2$.

In studying the third equation of \eqref{superalg}, let us note that
\begin{align}
\widetilde q^\beta{}_b|Q^\alpha{}_a\rangle
&=\Bigl\{-iq_{10}
\bigl[\eta
(\sinh\overline Y\widetilde W\epsilon)^{\alpha\beta}
-\frac{i}{2}\Tr\widetilde X^2
(\cosh\overline Y\epsilon)^{\alpha\beta}\bigr]
(\epsilon\cosh\overline Y{}')_{ba}\nonumber\\
&\quad-iq'_{01}
\bigl[\eta^*
(\epsilon\widetilde W{}'\sinh\overline Y{}')_{ba}
-\frac{i}{2}
\Tr\widetilde X{}'^2
(\epsilon\cosh\overline Y{}')_{ba}\bigr]
(\cosh\overline Y\epsilon)^{\alpha\beta}\nonumber\\
&\quad-i(q'_{01}-q_{10})
(\sinh\overline Y\widetilde X)^\alpha{}_b
(\widetilde X{}'\sinh\overline Y{}')^\beta{}_a\Bigr\}|V\rangle.
\end{align}
To satisfy the third equation, we have to first require that
\begin{align}
q_{10}=q'_{01}(=:q_1/2),
\label{mix}
\end{align}
for the cancellation of terms of odd powers in $Y$ and $Y'$.
Furthermore, using \eqref{eq:hYk} proved in appendix, we have the
formulas
\begin{align}
\bigl[\widetilde h,(\cosh\overline Y\epsilon)^{\alpha\beta}\bigr]
&=\eta(\sinh\overline Y\widetilde W\epsilon)^{\alpha\beta}
-\frac{i}{2}(\Tr Y\widetilde W)
(\cosh\overline Y\epsilon)^{\alpha\beta},\nonumber\\
\bigl[\widetilde h,(\epsilon\cosh\overline Y{}')_{ba}\bigr]
&=\eta^*(\epsilon\widetilde W{}'\sinh\overline Y{}')_{ba}
+\frac{i}{2}(\Tr\widetilde W{}'Y{}')
(\epsilon\cosh\overline Y{}')_{ba}.
\label{hcosh}
\end{align}
Hence, the third equation becomes
\begin{align}
q^\alpha{}_a|{\widetilde Q}^\beta{}_b\rangle
+{\widetilde q}^\beta{}_b|Q^\alpha{}_a\rangle
=\frac{i}{2}\bigl[\widetilde q_1h-q_1\widetilde h\bigr]
(\cosh\overline Y\epsilon)^{\alpha\beta}
(\epsilon\cosh\overline Y{}')_{ba}|V\rangle.
\label{eq:third}
\end{align}
This final result requires
\begin{align}
\widetilde q_1=q_1,
\end{align}
with the level matching condition $h=\widetilde h$.

Note that to study the third equation of \eqref{superalg}, the
dependence of the kinematical overlap on the worldsheet coordinate has
to be introduced in \cite{GSB,P}.
Here we have studied the same equation purely algebraically without
referring to the worldsheet coordinates.

\subsection{Chiral terms}
We can also consider the possibility of 
\begin{align}
|Q^\alpha{}_a\rangle&
=\Bigl\{\sum_{n,m}p_{nm}(Y^nXY'{}^m)^\alpha{}_a
+\sum_{n,m}p'_{mn}(Y^mX{}'Y'{}^n)^\alpha{}_a\Bigr\}
|V\rangle.
\label{eq:ansatz_p_nm}
\end{align}

The supersymmetry algebra puts constraints on the coefficients
$p_{12}=p_{32}=p'_{21}=p'_{23}=0$ as well as
\begin{align}
\frac{p_{10}}{i/2}=\frac{p'_{01}}{i/2}(=:p_1),\quad
\frac{p_{34}}{-2}=\frac{p'_{43}}{2}(=:p_7),\quad
\frac{p_{30}}{-2}=\frac{p'_{41}}{i/2}(=:p_>),\quad
\frac{p_{14}}{i/2}=\frac{p'_{03}}{2}(=:p_<).
\label{eq:p_gele}
\end{align}
Therefore, consistency with the first equation of \eqref{superalg}
requires
\begin{align}
|Q^\alpha{}_a\rangle
&=\Bigl\{p_1\bigl[\frac{i}{2}(YX)^\alpha{}_a+\frac{i}{2}(X'Y')^\alpha{}_a\bigr]
+p_7\bigl[-2(Y^3XY'^4)^\alpha{}_a+2(Y^4X'Y'^3)^\alpha{}_a\bigr]
\nonumber\\
&\quad+p_>\bigl[-2(Y^3X)^\alpha{}_a+\frac{i}{2}(Y^4X'Y')^\alpha{}_a\bigr]
+p_<\bigl[\frac{i}{2}(YXY'^4)^\alpha{}_a+2(X'Y'^3)^\alpha{}_a\bigr]
\Bigr\}
|V\rangle,
\nonumber\\
|H\rangle&=\Bigl\{p_1\frac{i}{2}\bigl(\Tr X^2+\Tr X'^2
-\Tr YW+\Tr W'Y'\bigr)
\nonumber\\
&\quad+\frac{p_7}{4}\bigl[\frac{i}{2}\Tr Y^4\Tr Y'^4
\bigl(\Tr X^2+\Tr X'^2\bigr)
+4\Tr Y^3W\Tr Y'^4+4\Tr Y^4\Tr W'Y'^3\bigr]
\nonumber\\
&\quad+\frac{p_>}{2}\bigl[\frac{i}{2}\Tr Y^4
\bigl(\Tr X^2+\Tr X'^2+\Tr W'Y'\bigr)
+4\Tr Y^3W\bigr]
\nonumber\\
&\quad+\frac{p_<}{2}\bigl[\frac{i}{2}\Tr Y'^4
\bigl(\Tr X^2+\Tr X'^2-\Tr YW\bigr)
+4\Tr W'Y'^3\bigr]
\Bigr\}|V\rangle.
\label{free}
\end{align}
Similarly $|\widetilde Q^\alpha{}_a\rangle$ and $|\widetilde H\rangle$
is determined from the second equation of \eqref{superalg}, while the
third equation holds identically with $|N\rangle=0$ if we require that
\begin{align}
p_1=\widetilde p_1,\quad p_7=\widetilde p_7,\quad
p_>=\widetilde p_>,\quad p_<=\widetilde p_<.
\end{align}
Furthermore, it is not difficult to notice that the result under this
ansatz can be summarized as
\begin{align}
|Q^\alpha{}_a\rangle=q^\alpha{}_a|W\rangle,\quad
|H\rangle=h|W\rangle,
\label{freeaction}
\end{align}
if we define the dressed kinematical overlap $|W\rangle$ as
\begin{align}
|W\rangle=\Bigl(p_1+\frac{p_>}{2}\Tr Y^4+\frac{p_<}{2}\Tr Y'^4
+\frac{p_7}{4}\Tr Y^4\Tr Y'^4\Bigr)|V\rangle.
\label{eq:Wket}
\end{align}

Previously, it was noticed \cite{dVPPRT,LR} that the free action on
the dressed kinematical overlap satisfies the supersymmetry algebra
as well.
However, there was no systematic analysis claiming that this is the
only alternative.
Here by listing all the possible terms with one $X$ or $X'$ inserted
\eqref{eq:ansatz_p_nm}, we find that all of the alternatives can be
written as the free action on the dressed kinematical overlap
$|W\rangle$.

\subsection{Supergravity limit}
Now let us study the supergravity limit.
In the supergravity limit, there is no difference between left-moving
modes and right-moving modes.
Therefore, the bosonic momenta $\widetilde X$ and $\widetilde X{}'$
should coincide with $X$ and $X'$, while the excited fermionic momenta
$W$, $W'$, $\widetilde W$ and $\widetilde W{}'$ without contribution
from zero modes should vanish identically.

It was further noticed \cite{GS} that in the supergravity Hamiltonian
on the flat spacetime, there appears an extra $\alg{u}(1)$ symmetry
coming from the duality algebra of the theory which can be expressed
as
\begin{align}
u=2-\frac{1}{2}\lambda^a\vartheta^a.
\end{align}
In the $\alg{so}(8)$-invariant formalism, the $\alg{u}(1)$ action on
various quantities is\footnote{The last relation can be understood
from the computation:
$\bigl(\sum_{r=1,2,3}u_{(r)}\bigr)\,
\delta^8(\lambda_{(1)}+\lambda_{(2)}+\lambda_{(3)})
=\bigl[2\times 3-(1/2)\times 8\bigr]\,
\delta^8(\lambda_{(1)}+\lambda_{(2)}+\lambda_{(3)})$.}
\begin{align}
[u,Y]=-\frac{1}{2}Y,\quad [u,X]=0,\quad u|V\rangle=2|V\rangle.
\end{align}
Therefore, the contribution other than
\begin{align}
|H\rangle\sim Y^4|V\rangle
\end{align}
should vanish identically in the supergravity limit.
In our current $\alg{so}(4)\times\alg{so}(4)$-invariant formalism, $Y$
and $Y'$ (or $X$ and $X'$) decomposed from the same quantity $Y$ (or
$X$) in the $\alg{so}(8)$ formalism should have the same charges.

Hereafter, let us determine some of the coefficients in \eqref{free}
by requiring the $\alg{u}(1)$ invariance in the supergravity
limit\footnote{Corresponding discussions were given previously in
\cite{LR}. It seems that their notation is different from ours.}.
Before starting it, let us first note that there are no $Y^2$ or $Y^6$
terms in \eqref{free}.
The condition of requiring that those terms in \eqref{exp} cancel
among themselves reduces to the same condition as \eqref{mix}.
Next, if we require the $Y^0$ (and $Y^8$) terms from \eqref{exp} and
from \eqref{free} cancel with each other, we find two equations for
$p_1$ (and $p_7$ respectively).
The two equations have the same solution only if \eqref{mix} holds and
the result is
\begin{align}
p_1=iq_1,\quad
p_7=\frac{-iq_1}{(4!)^2}.
\end{align}

\section{The $\alg{su}(2|2)$ light-cone string field theory}
After revisiting the pp-wave light-cone string field theory, it is now
clear to distinguish the property of general backgrounds with the
$\alg{su}(2|2)$ isometry from that special to the pp-wave
background.
In fact, in obtaining the most important result \eqref{exp} and
\eqref{free}, the main assumptions we have made is the last equation
of \eqref{CR} and \eqref{qV}.
By generalizing these equations, we hope that we can explore the
light-cone string field theory on general $\alg{su}(2|2)$
backgrounds.
Since we start our analysis from the fermionic momenta $Y$ and $Y'$
and act with the fermionic charges $q^\alpha{}_a$ to define other
quantities $X$, $X'$, $W$, $W'$, we can assign gradings and dimensions
to all of these building blocks from the outer automorphism of the
superalgebra $\alg{psu}(2|2)\ltimes{\mathbb R}^3$:
\begin{align}
\grd Y&=0,&\grd X=-\grd\widetilde X&=1/2,&\grd W=-\grd\widetilde W&=1,
\nonumber\\
\dim Y&=0,&\dim X=\dim\widetilde X&=1/2,&\dim W=\dim\widetilde W&=1.
\end{align}
We have to generalize the commutation relations \eqref{CR} respecting
the gradings\footnote{It is interesting to note that the grading
appeared as an extra symmetry in the $\alg{su}(2|2)$ spin chain model
\cite{spin} also plays an important role in constructing light-cone
string field theory.}, while the dimension serves in the role of the
expansion parameter.

The generalization of the right-hand side of the last equation in
\eqref{CR} can contain various terms including
\begin{align}
[q^\alpha{}_a,\widetilde X^b{}_{\dot b}]
=\sum_{m,n}y_{mn}(Y'^m){}^b{}_a(Y^n){}^\alpha{}_{\dot b}
+\cdots.
\label{eq:CRdeform}
\end{align}
The next corrections with bosonic momenta $X$, $X'$, $\widetilde X$
and $\widetilde X'$ inserted will contain about 100 terms.
Similarly, generalization of \eqref{qV} can be
\begin{align}
q^\alpha{}_a|V\rangle&
=\Bigl\{\sum_{n,m}v_{nm}(Y^nXY'^m)^\alpha{}_a
+\sum_{n,m}v'_{mn}(Y^mX'Y'^n)^\alpha{}_a\Bigr\}
|V\rangle,
\label{eq:qV_rs}
\end{align}
as well as terms obtained by replacing $X$ and $X'$ by $\widetilde X$
and $\widetilde X'$.

Our final direction will be to determine all of the coefficients so
that corrections to \eqref{CR} and \eqref{qV} are consistent with the
algebra \eqref{alg}.
For example, as previously in the case of the pp-wave background,
consistency with
\begin{align}
\{[q^\alpha{}_a,\widetilde X^c{}_{\dot c}],q^\beta{}_b\}
+\{q^\alpha{}_a,[q^\beta{}_b,\widetilde X^c{}_{\dot c}]\}
&=\epsilon^{\alpha\beta}\epsilon_{ab}[h,\widetilde X^c{}_{\dot c}],
\nonumber\\
\{q^\alpha{}_a,q^\beta{}_b\}|V\rangle
&=\epsilon^{\alpha\beta}\epsilon_{ab}h|V\rangle,
\nonumber\\
\{q^\alpha{}_a,\widetilde q^\beta{}_b\}|V\rangle
&=\epsilon^{\alpha\beta}\epsilon_{ab}n|V\rangle,
\label{eq:consistency_V}
\end{align}
gives strong constraints to the corrections.
In the third equation, we have imposed the condition
$r^a{}_b|V\rangle=l^{\alpha}{}_{\beta}|V\rangle=0$
because we assume that the invariance of the kinematical overlap
$|V\rangle$ under the $\alg{su}(2)\times\alg{su}(2)$ transformation
remains unmodified.

Due to a vast of possibilities, it is difficult to write down all of
the possible terms.
Instead, here we shall present some interesting possible corrections.

\subsection{Toy model I}
Let us consider a toy model for generalizations.
We assume the deformation of \eqref{eq:qV_rs} with $n=1,3$ and
$m=0,2,4$ instead of \eqref{qV}, keeping the commutation relations
\eqref{CR} unmodified.
{}From the consistency with the algebra on the overlap
\eqref{eq:consistency_V}, the coefficients of \eqref{eq:qV_rs} should
satisfy $v_{12}=v_{32}=v_{34}=v'_{21}=v'_{23}=v'_{43}=0$ and
\begin{align}
v_{30}=\widetilde v_{30},\quad
v_{34}=\widetilde v_{34},\quad
v'_{03}=\widetilde v'_{03},\quad
v'_{43}=\widetilde v'_{43},\quad
v_{34}+v'_{43}=0,
\label{eq:q6q6p_seigen}
\end{align}
in addition to $v_{10}=v'_{01}=i/2$.

By solving \eqref{superalg} with the ansatz \eqref{eq:onshell}, the
supersymmetry charges $|Q^{\alpha}{}_{a}\rangle$ and the Hamiltonian
$|H\rangle$ (\ref{exp}) are deformed as
\begin{align}
|Q^{\alpha}{}_{a}\rangle&=\Bigl\{
q\bigl[
(\sinh\overline Y)\widetilde X
\bigl(\cosh\overline Y{}'-\widehat v'Y{}'^4\bigr)
\bigr]^{\alpha}{}_a
+q'\bigl[
\bigl(\cosh\overline Y+\widehat vY{}^4\bigr)
\widetilde X{}'\sinh\overline Y{}'
\bigr]^{\alpha}{}_a
\Bigr\}|V\rangle,\nonumber\\
|H\rangle&=\Bigl\{
b\bigl[
\Tr Y{}^4(1-(\widehat v'/2)\Tr Y{}'^4)
-\Tr Y{}'^4(1+(\widehat v/2)\Tr Y{}^4)
\bigr]
\nonumber\\
&\quad+q\bigl[
\eta\Tr X
(\cosh{\overline Y}-\widehat vY{}^4)
{\widetilde X}
(\cosh{\overline Y}{}'-\widehat v'Y{}'^4)
-\eta\widehat b{\rm Tr}X Y{}^4\widetilde X Y{}'^4
\nonumber\\&\qquad\qquad
+\eta^*\Tr\sinh{\overline Y}{\widetilde X}
\sinh{\overline Y}{}'X{}'
\bigr]
\nonumber\\
&\quad+q'\bigr[
\eta^*\Tr(\cosh{\overline Y}+\widehat vY{}^4)
{\widetilde X}'
(\cosh{\overline Y}{}'+\widehat v'Y{}'^4)
X{}'
-\eta^*\widehat b\Tr Y{}^4\widetilde X{}'Y{}'^4 X{}'
\nonumber\\&\qquad\qquad
+\eta\Tr X\sinh{\overline Y}
{\widetilde X}{}'\sinh{\overline Y}{}'
\bigr]
\Bigr\}|V\rangle,
\end{align}
where we have defined $\widehat v=v_{30}/2$, $\widehat v'=v'_{03}/2$
and $\widehat b=v_{34}/4=-v'_{43}/4$.
In this case, \eqref{eq:third} becomes
\begin{align}
q^\alpha{}_a|{\widetilde Q}^\beta{}_b\rangle
+{\widetilde q}^\beta{}_b|Q^\alpha{}_a\rangle
=\frac{i}{2}q_1\bigl[h-\widetilde h\bigr]
\bigl[(\cosh\overline Y\epsilon)^{\alpha\beta}
(\epsilon\cosh\overline Y{}')_{ba}
+4\widehat b(Y{}^4\epsilon)^{\alpha\beta}
(\epsilon Y{}'^4)_{ba}\bigr]
|V\rangle.
\end{align}

On the other hand, if we take the ansatz of the form
\eqref{eq:ansatz_p_nm}, \eqref{superalg} is solved by
\begin{align}
{}|Q^\alpha{}_a\rangle
&=\Bigl\{\frac{i}{2}p_1\bigl[(YX)^\alpha{}_a+(X'Y')^\alpha{}_a\bigr]
\nonumber\\
&\quad+(-2p_7+v_{30}p_{<}+4v_{34}p_1)(Y^3XY'^4)^\alpha{}_a
+(2p_7+v'_{03}p_{>}+4v'_{43}p_1)(Y^4X'Y'^3)^\alpha{}_a
\nonumber\\
&\quad+(-2p_{>}+v_{30}p_1)(Y^3X)^\alpha{}_a
+\frac{i}{2}p_>(Y^4X'Y')^\alpha{}_a
\nonumber\\
&\quad+\frac{i}{2}p_<(YXY'^4)^\alpha{}_a
+(2p_{<}+v'_{03}p_1)(X'Y'^3)^\alpha{}_a
\Bigr\}
|V\rangle,
\end{align}
where we have redefined $p_1$, $p_7$, $p_>$ and $p_<$ by
\begin{align}
&\frac{p_{10}}{i/2}=\frac{p'_{01}}{i/2}=:p_1,\quad
\frac{p_{34}}{-2}+\frac{v_{30}}{2}p_{<}+2v_{34}p_1
=\frac{p'_{43}}{2}-\frac{v'_{03}}{2}p_{>}-2v'_{43}p_1
=:p_7,\nonumber\\
&\frac{p_{30}}{-2}+\frac{v_{30}}{2}p_1=\frac{p'_{41}}{i/2}=:p_{>},
\quad 
\frac{p_{14}}{i/2}=\frac{p'_{03}}{2}-\frac{v'_{03}}{2}p_1=:p_{<},
\end{align}
instead of \eqref{eq:p_gele}.
We find that the above expression can also be rewritten as the action
of free charges on a dressed kinematical overlap \eqref{freeaction}
with the dressed kinematical overlap formally taking the same
expression as \eqref{eq:Wket}.

\subsection{Toy model II}

As a next example, let us consider a type of deformations of the last
equation of \eqref{CR} with $q^{\alpha}{}_a|V\rangle$ \eqref{qV}
unchanged.
We introduce the coefficient of $Y^3$, namely $y_{03}$ in
\eqref{eq:CRdeform}, and a counterpart in
$[q^{\alpha}{}_a,\widetilde X'^{\beta}{}_{\dot\beta}]$, which we
denote as $y'_{30}$.
In this case, the Jacobi identities such as the first equation of
\eqref{eq:consistency_V} are broken.
To solve this problem, further deformation terms in the commutation
relations are required.
Actually, we find a consistent deformation of commutation relations,
up to terms of dim 1,
\begin{align}
[q^\alpha{}_a,\widetilde X^{\dot b}{}_b]
&=-\frac{i}{2}(Y\epsilon)^{\dot b\alpha}\epsilon_{ab}
+y_{03}\bigl[(Y^3\epsilon)^{\dot b\alpha}\epsilon_{ab}
+2i(Y\epsilon)^{\dot b\alpha}(\epsilon X\widetilde X)_{ab}
+4i(YX)^{\alpha}{}_a\widetilde X^{\dot b}{}_b\bigr],\nonumber\\
[q^\alpha{}_a,\widetilde X'^\beta{}_{\dot\beta}]
&=\frac{i}{2}\epsilon^{\beta\alpha}(\epsilon Y')_{a\dot\beta}
-y'_{30}\bigl[\epsilon^{\beta\alpha}(\epsilon Y'^3)_{a\dot\beta}
+2i(\widetilde X'X'\epsilon)^{\beta\alpha}(\epsilon Y')_{a\dot\beta}
+4i(X'Y')^{\alpha}{}_a\widetilde X'^{\beta}{}_{\dot\beta}\bigr],
\end{align}
and their counterparts of $[\widetilde q^\alpha{}_a, X^{\dot b}{}_b]$
and $[\widetilde q^\alpha{}_a,X'^\beta{}_{\dot\beta}]$, which satisfy
the first equation\footnote{We note that the Jacobi identity composed
of $q^{\alpha}{}_a,\widetilde q^{\alpha}{}_a$ and 
$\widetilde X^c{}_{\dot c}$ does not give any constraints because
$\{q^{\alpha}{}_a,\widetilde W^{\beta}{}_{\dot b}\}$ is not defined
yet.} of \eqref{eq:consistency_V} up to terms of dim 1/2.
From the consistency with the third equation of
\eqref{eq:consistency_V},
\begin{align}
y_{03}=\widetilde y_{03},\quad
y'_{30}=\widetilde y'_{30},
\end{align}
are required.

With the above deformation and the ansatz \eqref{eq:onshell}, the
supersymmetry charges are obtained by the same formula as the first
equation of \eqref{exp}, which satisfy the first algebraic relation of
\eqref{superalg} up to terms of dim 1.
Although the relation \eqref{eq:third} is unchanged, the Hamiltonian
$|H\rangle$ is deformed from the second equation of \eqref{exp} by
\begin{align}
\Bigl[
\eta y_{03}q\Tr Y{}^4\Bigl(1-\frac{1}{48}\Tr Y{}'^4\Bigr)
+\eta^*y'_{30}q'\Tr Y{}'^4\Bigl(1-\frac{1}{48}\Tr Y{}^4\Bigr)
\Bigr]|V\rangle.
\end{align}

\section{Discussions}

We have revisited the construction of light-cone string field theory
on the pp-wave background and clarified its algebraic structure.
Among others, we find the centrally extended subalgebra
$\alg{psu}(2|2)\ltimes{\mathbb R}^3$ of the whole pp-wave isometry
plays an essential role in the construction.
In the meanwhile, we distinguish the properties satisfied by general
$\alg{su}(2|2)$ backgrounds from those special to the pp-wave
background.
We also make several improvements for the construction on the pp-wave
backgrounds:
\begin{itemize}
\item
Our systematic analysis reproduces the famous previous results
\cite{P} in a succinct form \eqref{exp}.
\item
Previously, a dependence of $|V\rangle$ on the worldsheet coordinate
had to be introduced in \cite{GSB,P} to study the orthogonality
between two supercharges ${\mathcal Q}^\alpha{}_a$ and 
$\widetilde{\mathcal Q}^\alpha{}_a$.
Here we show that the same relation can be derived purely
algebraically.
\item
We present a thorough study of possible contributions from the chiral
terms.
We find that all of the interaction terms $|J\rangle$ can be put into
the forms $|J\rangle=j|W\rangle$ \eqref{freeaction} where free charges
$j$ are acting on a dressed kinematical overlap $|W\rangle$.
\item
We determine some of the coefficients of the chiral terms from the
abelian duality group in the supergravity limit.
\end{itemize}

After the clarification, we propose a strategy for extending the
background from the pp-wave one to a more general one with the
isometry $\alg{su}(2|2)$ and present some examples of
generalizations.
As further directions, we hope to identify some of these
generalizations as $AdS_5\times S^5$, and investigate AdS/CFT
correspondence with the string interactions.

\section*{Acknowledgement}
We would like to thank J.~Gomis, H.~Hata, Y.~Kazama, H.~Kunitomo,
S.~Lee and S.~Teraguchi for valuable discussions and especially
S.~Dobashi for collaborations at the very early stage of the present
work.
We are grateful to the organizers of Summer Institute 2009 at
Fuji-Yoshida for providing a stimulating environment and S.M.\ is
grateful to Benasque and RIKEN for hospitality where various stages of
the current work were done.
The work of I.K.\ is supported partly by JSPS Research Grant for the
Trend Survey of Scientific Researches and JSPS Grant-in-Aid for
Scientific Research (C) [\#21540269] and it was supported partly by a
Special Postdoctoral Researchers Program at RIKEN.
The work of S.M.\ is supported partly by Grant-in-Aid from Daiko
Foundation and partly by Grant-in-Aid for Young Scientists (B)
[\#21740176] from the Japan Ministry of Education, Culture, Sports,
Science and Technology.

\appendix

\section{Some useful formulas}
In this appendix we shall prove some useful formulas including
\eqref{qnm}--\eqref{XYmn}.
For Grassmann odd quantities $Y^{\alpha}{}_{\dot a}$
($\alpha=1,2,a=1,2$), the following identities hold:
\begin{align}
Y^{\alpha}{}_{\dot a}Y^{\beta}{}_{\dot b}
&=-\frac{1}{2}(Y^2\epsilon)^{\alpha\beta}\epsilon_{\dot a\dot b}
-\frac{1}{2}(\epsilon Y^2)_{\dot a\dot b}\epsilon^{\alpha\beta},
\nonumber\\
Y^{\alpha}{}_{\dot a}Y^{\beta}{}_{\dot b}Y^{\gamma}{}_{\dot c}
&=\frac{1}{9}(Y^3)^{\delta}{}_{\dot d}
\bigl(\delta^{\alpha}_{\delta}\epsilon^{\beta\gamma}
(\delta^{\dot d}_{\dot c}\epsilon_{\dot a\dot b}
-\delta^{\dot d}_{\dot b}\epsilon_{\dot c\dot a})
+\delta^{\beta}_{\delta}\epsilon^{\gamma\alpha}
(\delta^{\dot d}_{\dot a}\epsilon_{\dot b\dot c}
-\delta^{\dot d}_{\dot c}\epsilon_{\dot a\dot b})
+\delta^{\gamma}_{\delta}\epsilon^{\alpha\beta}
(\delta^{\dot d}_{\dot b}\epsilon_{\dot c\dot a}
-\delta^{\dot d}_{\dot a}\epsilon_{\dot b\dot c})
\bigr),
\nonumber\\
Y^{\alpha}{}_{\dot a}Y^{\beta}{}_{\dot b}Y^{\gamma}{}_{\dot c}Y^{\delta}{}_{\dot d}
&=-\frac{1}{36}\Tr Y^4\nonumber\\
&\hspace{-1cm}\times\bigl(
\epsilon^{\delta\alpha}\epsilon^{\beta\gamma}
(\epsilon_{\dot d\dot c}\epsilon_{\dot a\dot b}
-\epsilon_{\dot d\dot b}\epsilon_{\dot c\dot a})
+\epsilon^{\delta\beta}\epsilon^{\gamma\alpha}
(\epsilon_{\dot d\dot a}\epsilon_{\dot b\dot c}
-\epsilon_{\dot d\dot c}\epsilon_{\dot a\dot b})
+\epsilon^{\delta\gamma}\epsilon^{\alpha\beta}
(\epsilon_{\dot d\dot b}\epsilon_{\dot c\dot a}
-\epsilon_{\dot d\dot a}\epsilon_{\dot b\dot c})
\bigr),
\label{Yprod}
\end{align}
where
\begin{align}
&(Y^2\epsilon)^{\alpha\beta}
=Y^{\alpha}{}_{\dot a}Y^{\dot a}{}_{\gamma}\epsilon^{\gamma\beta}
=Y^{\alpha}{}_{\dot a}(\epsilon_{\gamma\delta}
Y^{\delta}{}_{\dot b}\epsilon^{\dot b\dot a})\epsilon^{\gamma\beta}
=(Y^2\epsilon)^{\beta\alpha},\nonumber\\
&(\epsilon Y^2)_{\dot a\dot b}
=\epsilon_{\dot a\dot c}Y^{\dot c}{}_{\alpha}Y^{\alpha}{}_{\dot b}
=\epsilon_{\dot a\dot c}(\epsilon_{\alpha\beta}
Y^{\beta}{}_{\dot d}\epsilon^{\dot d\dot c})Y^{\alpha}{}_{\dot b}
=(\epsilon Y^2)_{\dot b\dot a}.
\end{align}

Let us consider commutation relations between
$\Tr\xi q\equiv\xi^a{}_{\alpha}q^{\alpha}{}_{a}$ and $Y$, $Y'$, where
Grassmann odd parameters $\xi^a{}_{\alpha}$ are introduced for
simplicity of indices.
The results are given by
\begin{align}
[\Tr\xi q,Y^{\alpha}{}_{\dot a}]
&=-(\xi X)^{\alpha}{}_{\dot a},&
[\Tr\xi q,\bar Y^{\dot a}{}_{\alpha}]
&=(\bar X\xi)^{\dot a}{}_{\alpha},
\nonumber\\
[\Tr\xi q,Y'^a{}_{\dot\alpha}]
&=(\xi X')^a{}_{\dot\alpha},&
[\Tr\xi q,\bar Y'^{\dot\alpha}{}_{a}]
&=-(\bar X'\bar\xi)^{\dot\alpha}{}_{a},
\label{eq:xiq_com4}
\end{align}
with\footnote{We attach bars ($\bar{~}$) to avoid confusions in the
following matrix notation.
These bars should not be confused with the overlines of $\overline{Y}$
and $\overline{Y}{}'$.}
$\bar\xi^{\alpha}{}_{a}
=\epsilon_{ab}\xi^b{}_{\beta}\epsilon^{\beta\alpha}$,
$\bar Y^{\dot a}{}_{\alpha}
=\epsilon_{\alpha\beta}Y^{\beta}{}_{\dot b}\epsilon^{\dot b\dot a}$,
$\bar Y'^{\dot\alpha}{}_{a}
=\epsilon_{ab}Y^b{}_{\dot\beta}\epsilon^{\dot\beta\dot\alpha}$ as well as
$\bar X^{\dot a}{}_a
=\epsilon_{ab}X^b{}_{\dot b}\epsilon^{\dot b\dot a}$,
$\bar X'^{\dot\alpha}{}_{\alpha}
=\epsilon_{\alpha\beta}X'^{\beta}{}_{\dot\beta}\epsilon^{\dot\beta\dot\alpha}$.
If we define $4\times 4$ matrices ${\bs Y},{\bs Y'}$ as
\begin{align}
{\bs Y}=({\bs Y}_{(\alpha,\dot a),(\beta,\dot b)})=
\begin{pmatrix}
0&Y^{\alpha}{}_{\dot b}\\
\bar Y^{\dot a}{}_{\beta}&0
\end{pmatrix},\quad
{\bs Y}'=({\bs Y}'_{(a,\dot\alpha),(b,\dot\beta)})=
\begin{pmatrix}
0&Y'^a{}_{\dot\beta}\\
\bar Y'^{\dot\alpha}{}_{b}&0
\end{pmatrix},
\end{align}
\eqref{eq:xiq_com4} can be rewritten as
\begin{align}
[\Tr\xi q,{\bs Y}]={\bs X}_\xi,\quad
[\Tr\xi q,{\bs Y}']={\bs X}'_\xi,
\label{eq:TrxiqYY'}
\end{align}
with ${\bs X}_\xi$ and ${\bs X}'_\xi$ defined by
\begin{align}
{\bs X}_\xi=
\begin{pmatrix}
0&-(\bar\xi X)^{\alpha}{}_{\dot b}\\
(\bar X\xi)^{\dot a}{}_{\beta}&0
\end{pmatrix},\quad
{\bs X}'_\xi=
\begin{pmatrix}
0&(\xi X')^a{}_{\dot \beta}\\
-(\bar X'\bar\xi)^{\dot \alpha}{}_b&0
\end{pmatrix}.
\label{eq:bsxx''}
\end{align}
Using the relations 
\begin{align}
{\bs Y}{\bs X}_\xi{\bs Y}
=\frac{1}{2}({\bs X}_\xi{\bs Y}^2+{\bs Y}^2{\bs X}_\xi),\quad
{\bs Y}'{\bs X}'_\xi{\bs Y}'
=\frac{1}{2}({\bs X}'_\xi{\bs Y}'^2+{\bs Y}'^2{\bs X}'_\xi),
\end{align}
which follow from \eqref{Yprod}, the commutation relations
\eqref{eq:TrxiqYY'} can be easily generalized into
\begin{align}
[\Tr\xi q,{\bs Y}^k]
=\frac{k}{2}({\bs X}_\xi{\bs Y}^{k-1}+{\bs Y}^{k-1}{\bs X}_\xi),\quad
[\Tr\xi q,{\bs Y}'^k]
=\frac{k}{2}({\bs X}'_\xi{\bs Y}'^{k-1}+{\bs Y}'^{k-1}{\bs X}'_\xi),
\label{eq:TrxiqYk}
\end{align}
for non-negative integers $k$.
More explicitly, they can be rewritten as
\begin{align}
[\Tr\xi q,{\bs Y}^n]
&=\frac{n}{2}
\begin{pmatrix}
0&-\bar\xi X\bar Y^{n-1}-Y^{n-1}\bar\xi X\\
\bar X\xi Y^{n-1}+\bar Y^{n-1}\bar X\xi&0
\end{pmatrix},\nonumber\\
[\Tr\xi q,{\bs Y}'^n]
&=\frac{n}{2}
\begin{pmatrix}
0&\xi X'\bar Y'^{n-1}+Y'^{n-1}\xi X'\\
-\bar X'\bar\xi Y'^{n-1}-\bar Y'^{n-1}\bar X'\bar\xi&0
\end{pmatrix},
\label{eq:xiqbfYpn}
\end{align}
for $n=1,3$, and 
\begin{align}
[\Tr\xi q,{\bs Y}^m]
&=\frac{m}{2}
\begin{pmatrix}
-\bar\xi X\bar Y^{m-1}+Y^{m-1}\bar X\xi&0\\
0&\bar X\xi Y^{m-1}-\bar Y^{m-1}\bar\xi X
\end{pmatrix},
\nonumber\\
[\Tr\xi q,{\bs Y}'^m]
&=\frac{m}{2}
\begin{pmatrix}
\xi X'\bar Y'^{m-1}-Y'^{m-1}\bar X'\bar\xi&0\\
0&-\bar X'\bar\xi Y'^{m-1}+\bar Y'^{m-1}\xi X'
\end{pmatrix},
\label{eq:xiqbfYpm}
\end{align}
for $m=0,2,4$, where we denote the matrix products of $Y$ and $Y'$ as
\begin{align}
Y^k=Y\bar Y Y\cdots,\quad\bar Y^k=\bar Y Y\bar Y\cdots,\quad
Y'^k=Y'\bar Y' Y'\cdots,\quad\bar Y'^k=\bar Y' Y'\bar Y'\cdots.
\end{align}
By removing $\xi^a{}_{\alpha}$ from the above formulas
\eqref{eq:xiqbfYpn} and \eqref{eq:xiqbfYpm}, we obtain
\begin{align}
\{q^{\alpha}{}_{a},(Y^n)^{\beta}{}_{\dot c}\}
&=-\frac{n}{2}\left[
\epsilon^{\beta\alpha}(\epsilon X\bar Y^{n-1})_{a\dot c}
+(Y^{n-1}\epsilon)^{\beta\alpha}(\epsilon X)_{a\dot c}
\right],\nonumber\\
{}[q^{\alpha}{}_{a},(Y^m)^{\beta}{}_{\gamma}]
&=-\frac{m}{2}\left[
\epsilon^{\beta\alpha}(\epsilon X\bar Y^{m-1})_{a\gamma}
+(Y^{m-1}\bar X)^{\beta}{}_{a}\delta^{\alpha}_{\gamma}
\right],\nonumber\\
\{q^{\alpha}{}_{a},(\bar Y'^n)^{\dot\gamma}{}_{b}\}
&=-\frac{n}{2}\left[
(\bar X'\epsilon)^{\dot\gamma\alpha}(\epsilon\bar Y'^{n-1})_{ab}
+(\bar Y'^{n-1}\bar X'\epsilon)^{\dot\gamma\alpha}\epsilon_{ab}
\right],\nonumber\\
{}[q^{\alpha}{}_{a},(Y'^m)^c{}_{b}]
&=\frac{m}{2}\left[
\delta^c_a(X'\bar Y'^{m-1})^{\alpha}{}_{b}
+(Y'^{m-1}\bar X'\epsilon)^{c\alpha}\epsilon_{ab}
\right],
\end{align}
where $n=1,3$ and $m=0,2,4$.
These formulas are essentially \eqref{qnm} and \eqref{qmn}.

Next we shall turn to the derivation of the remaining formulas.
Noting $\epsilon_{\alpha\beta}\delta_{\gamma}^{\delta}+
\epsilon_{\beta\gamma}\delta_{\alpha}^{\delta}+
\epsilon_{\gamma\alpha}\delta_{\beta}^{\delta}=0$
and similar identities, we find
\begin{align}
{\bf 1}_4\Tr(\xi Y\bar X)
={\bs X}_\xi{\bs Y}-{\bs Y}{\bs X}_\xi,\quad
{\bf 1}_4\Tr(\xi X'\bar Y')
={\bs X}'_\xi{\bs Y}'-{\bs Y}'{\bs X}'_\xi,
\end{align}
where ${\bf 1}_4$ is the identity matrix of $4\times 4$.
Using the above formulas repeatedly, we obtain
\begin{align}
{\bs Y}^{k-1}\Tr(\xi Y\bar X)
=\frac{1}{k}({\bs X}_\xi{\bs Y}^k-{\bs Y}^k{\bs X}_\xi),\quad
{\bs Y}'^{k-1}\Tr(\xi X'\bar Y')
=\frac{1}{k}({\bs X}'_\xi{\bs Y}'^k-{\bs Y}'^k{\bs X}'_\xi),
\end{align}
with $k$ being a positive integer.
These relations are nothing but \eqref{YXnm}-\eqref{XYnm} and
\eqref{YXmn}-\eqref{XYmn}.

Similarly, noting \eqref{eq:hY=W} or
\begin{align}
[h,{\bs Y}]={\bs W},\quad
[h,{\bs Y}']={\bs W}',
\end{align}
where
\begin{align}
{\bs W}=
\begin{pmatrix}
0&W^{\alpha}{}_{\dot b}\\
\bar W^{\dot a}{}_{\beta}&0
\end{pmatrix},\quad
{\bs W}'=
\begin{pmatrix}
0&W'^a{}_{\dot\beta}\\
\bar W'^{\dot\alpha}{}_b&0
\end{pmatrix},
\end{align}
we find the multiplication formulas for ${\bs Y},{\bs Y}',{\bs W}$ and
${\bs W}'$
\begin{align}
{\bs Y}{\bs W}{\bs Y}
&=\frac{1}{2}({\bs W}{\bs Y}^2+{\bs Y}^2{\bs W}),&
{\bs Y}'{\bs W}'{\bs Y}'
&=\frac{1}{2}({\bs W}'{\bs Y}'^2+{\bs Y}'^2{\bs W}'),
\nonumber\\
{\bs Y}^k{\bs W}
&={\bs W}{\bs Y}^k+k{\bs Y}^{k-1}\Tr(Y\bar W),&
{\bs Y}'^k{\bs W}'
&={\bs W}'{\bs Y}'^k-k{\bs Y}'^{k-1}\Tr(W'\bar Y'),
\end{align}
as well as the $h$ action,
\begin{align}
[h,{\bs Y}^k]
&=k{\bs W}{\bs Y}^{k-1}
+\frac{k(k-1)}{2}{\bs Y}^{k-2}\Tr(Y\bar W)
=k{\bs Y}^{k-1}{\bs W}-\frac{k(k-1)}{2}{\bs Y}^{k-2}
\Tr(Y\bar W),
\nonumber\\
[h,{\bs Y}'^k]
&=k{\bs W}'{\bs Y}'^{k-1}
-\frac{k(k-1)}{2}{\bs Y}'^{k-2}\Tr(W'\bar Y')
=k{\bs Y}'^{k-1}{\bs W}'
+\frac{k(k-1)}{2}{\bs Y}'^{k-2}\Tr(W'\bar Y'),
\label{eq:hYk}
\end{align}
where $k$ is a positive integer.
This final result is helpful in studying \eqref{hcosh}.

\end{document}